\documentclass[10pt,conference]{IEEEtran}
\IEEEoverridecommandlockouts
\usepackage[noadjust]{cite}
\usepackage{amsmath}
\usepackage{filecontents}
\newcommand{\SocketAIScanner}{SocketAI  }
\newcommand{\SocketAIScannernogap}{SocketAI}
\newcommand{\totalfile}{20,926 }
\newcommand{\totaluniquefile}{18,754 }
\newcommand{\sastfile}{4,146 }
\newcommand{\sastpackage}{2,861 }
\newcommand{\malwareversion}{2,776 }
\newcommand{\malwarepackage}{2,180 }

\newcommand{\neutralversion}{2,974 }
\newcommand{\neutralpackage}{2,935 }

\newcommand{\uniquepackage}{5,115 }
\newcommand{\socket}{Socket }

\usepackage[bookmarks=false]{hyperref}

\newcommand{\realsearchgoal}{\textit{ The goal of this study is to aid security analysts in detecting malicious packages by empirically studying the effectiveness of Large Language Models (LLMs) in detecting malicious code.}}

\newcommand{\RQOne}{What performance can be achieved in malicious code detection using LLMs compared with static analysis (baseline comparison), in terms of precision, recall, and F1 score? }
\newcommand{\RQTwo}{What is the effectiveness of using a static analyzer as a pre-screener with \SocketAIScanner in terms of the number of files that need to be analyzed and the associated cost?} 
\newcommand{\RQThree}{What types of malicious packages are detected or missed by LLMs? }

\usepackage{multirow}
\usepackage{array}
\usepackage{caption}
\usepackage{tcolorbox}
\usepackage{xcolor}
\definecolor{lightgray}{gray}{0.9}
\makeatletter
\usepackage{longtable}
\usepackage{listings}
\definecolor{codegreen}{rgb}{0,0.6,0}
\definecolor{codegray}{rgb}{0.5,0.5,0.5}
\definecolor{codepurple}{rgb}{0.58,0,0.82}
\definecolor{backcolour}{rgb}{0.95,0.95,0.92}

\lstdefinelanguage{JavaScript}{
  keywords={break, case, catch, continue, debugger, default, delete, do, else, finally, for, function, if, in, instanceof, return, switch, this, throw, try, typeof, var, let,  while, with},
  morecomment=[l]{//},
  morecomment=[s]{/*}{*/},
  morestring=[b]',
  morestring=[b]",
  ndkeywords={class, export, boolean, throw, implements, import, this},
  keywordstyle=\color{blue}\bfseries,
  ndkeywordstyle=\color{darkgray}\bfseries,
  identifierstyle=\color{black},
  commentstyle=\color{purple}\ttfamily,
  stringstyle=\color{red}\ttfamily,
  sensitive=true
}
\makeatother

\author{\IEEEauthorblockN{Nusrat Zahan\IEEEauthorrefmark{1}, Philipp Burckhardt\IEEEauthorrefmark{2}, Mikola Lysenko\IEEEauthorrefmark{2}, Feross Aboukhadijeh\IEEEauthorrefmark{2}, Laurie Williams\IEEEauthorrefmark{1}}
\IEEEauthorblockA{\IEEEauthorrefmark{1}North Carolina State University \\
Email: [nzahan, lawilli3]@ncsu.edu}
\IEEEauthorblockA{\IEEEauthorrefmark{2}Socket, Inc \\
Email: [philipp, mik, feross]@socket.dev}}
\begin{document}
\date{}

\title{\Large \bf {Leveraging Large Language Models to Detect npm Malicious Packages}}

\maketitle

\begin{abstract}
Existing malicious code detection techniques %
demand the integration of multiple tools to detect different malware patterns, often suffering from high misclassification rates. Therefore, malicious code detection techniques could be enhanced by adopting advanced, more automated approaches to achieve high accuracy and a low misclassification rate. \realsearchgoal ~We present \SocketAIScannernogap, a malicious code review workflow to detect malicious code. %
To evaluate the effectiveness \SocketAIScannernogap, we leverage a benchmark dataset of \uniquepackage npm packages, of which \malwarepackage packages have malicious code. We conducted a baseline comparison of GPT-3 and GPT-4 models with the state-of-the-art CodeQL static analysis tool, using 39 custom CodeQL rules developed in prior research to detect malicious Javascript code. We also compare the effectiveness of static analysis as a pre-screener with \SocketAIScanner workflow, measuring the number of files that need to be analyzed and the associated costs. Additionally, we performed a qualitative study to understand the types of malicious packages detected or missed by our workflow. Our baseline comparison demonstrates a 16\%  and 9\% improvement over static analysis in precision and F1 scores, respectively. GPT-4 achieves higher accuracy with 99\% precision and 97\% F1 scores, while GPT-3 offers a more cost-effective balance at 91\% precision and 94\% F1 scores. Pre-screening files with a static analyzer reduces the number of files requiring LLM analysis by 77.9\% and decreases costs by 60.9\% for GPT-3 and 76.1\% for GPT-4. Our qualitative analysis identified data theft, execution of arbitrary code, and suspicious domain categories as the top detected malicious packages. %

\end{abstract}

\section{Introduction}

Open-source package repositories are often misused for software supply chain attacks~\cite{ladisa2022risk}. Attackers attempt to execute supply chain attacks by injecting malicious code into existing packages or by creating new infected packages and tricking users into downloading them ~\cite{ladisa2022risk,ohm2020backstabber}. %
Classifying a package as malicious is challenging because malicious code detection necessitates comprehending the reasoning about the possible purpose and intent of the code. Given the increasing rate of software supply chain attacks, the demand for approaches to detect such attacks is also increasing.

The current malicious package detection~\cite{garrett2019detecting, gonzalez2021anomalicious, zahan2022weak, ohm2022towards, scalco2022feasibility, sejfia2022practical, vu2020towards, vu2023bad, wang2021feasibility} including static, rules, heuristics, differential analysis, and different machine learning techniques encounter challenges of incomplete solutions, high misclassification rates~\cite{ohm2023sok}. Additionally, no single technique can comprehensively address different attack types, like detecting malicious code injection, dependency confusion, and typosquat, and often fails to support languages like TypeScript and CoffeeScript or research packages~\cite{ohm2023sok}. As a result, practitioners need to integrate multiple tools to detect malware patterns. An advanced detection technique is needed to identify diverse attack types effectively and minimize manual review efforts with minimal misclassification. LLMs have the potential to address these limitations by detecting diverse attacks through their advanced understanding of context, semantics, and patterns, thereby reducing reliance on multiple tools.

While the existing literature lacks the contribution of using LLMs to detect malicious code, organizations are adopting LLMs as an advanced technique for threat analysis~\cite{elasitic_AI,Socket_AI_article}. However, assessing the performance of LLMs in production presents challenges: 1) comprehensive evaluation in a production environment is resource-intensive: 2) LLM missing actual malicious code (false negative); 3) validation due to the lack of ground truth; and 4) after reporting potential malicious packages to the package registry, the package managers often only acknowledge receipt of the reports without confirming the presence of malicious code. Hence, we conduct an empirical study to evaluate LLMs' performance in malicious code detection.  
\realsearchgoal~In this study, we address the following research questions~(RQs)\textemdash
\begin{itemize}
    \item \textbf{RQ1}: \RQOne
    \item \textbf{RQ2}: \RQTwo
    \item \textbf{RQ3}: \RQThree
\end{itemize}

To the best of our knowledge \SocketAIScannernogap, a malicious code review workflow is a first-ever study leveraging LLMs to detect malicious code and compares the effectiveness of the workflow with static analysis tools.  The workflow leverages Iterative Self-Refinement~\cite{madaan2023self}, LLM-as-a-Judge~\cite{zheng2023judging} and Zero-Shot-Role-Play-Chain of Thought (CoT) prompting~\cite{kojima2205large, kong2023better} techniques in ChatGPT models (GPT-3 and GPT-4).  We leveraged MalwareBench~\cite{zahan2024malwarebench}, a benchmark dataset, and studied \uniquepackage npm packages, including \malwarepackage malicious and \neutralpackage neutral packages, to evaluate the performance of LLMs. %

To answer RQ1, The paper presents a baseline comparison between ChatGPT models and the mature, state-of-the-art commercial tool, CodeQL. We used 39 custom rules of CodeQL that have been used in prior study~\cite{froh2023differential} to detect malicious Javascript packages. We compare results using the standard machine learning (ML) evaluation metrics, precision, recall, and F1 score. We answer RQ2 by comparing the effectiveness of using ``\SocketAIScanner only'' versus using ``CodeQL as a pre-screener with \SocketAIScanner'' based on the number of files analyzed and the associated costs. For RQ3, we provide qualitative insights by manually evaluating the model’s responses to understand the types of malicious packages identified or missed by ChatGPT and demonstrate the opportunities and strengths of these models. For the benefit of other researchers, we further report issues we encountered while using ChatGPT.  Our contributions are:

\begin{itemize}
    \item \SocketAIScannernogap, a malicious code review workflow to detect malicious packages in the npm ecosystem; 
    \item Malware-domain-specific prompt to trigger LLMs reasoning in detecting malicious code;
    \item A baseline comparison of \SocketAIScanner with CodeQL tool to understand the effectiveness of LLMs; and 
    \item Strengths and shortcomings of ChatGPT for large-scale code review.
\end{itemize}

This paper is organized as follows: Section \ref{approach} includes our design approach, discussing our dataset, allocated cost,  package selection, and model selection criteria. Section \ref{SocketAIScanner} covers our proposed workflow of \SocketAIScannernogap.  We answer our three RQs in section \ref{RQ1}, section \ref{RQ2}, and section \ref{RQ3}. Then, we discussed challenges in using LLM in section \ref{Challange}. The related work section is covered in section \ref{Related_work}. We close with limitations and discussion of our work~(Section \ref{limitation} and \ref{conclusion}).

\section{\textbf{Dataset and LLM Selection}} \label{approach}
Section \ref{malwarebench} discusses the benchmark dataset used in this study, Section \ref{LLmodel} discusses LLMs selection process, and Section \ref{package_selection} explains our package selection criteria with cost constraint discussed in Section ~\ref{LLM_Cost}. 

\subsection{\textbf{Benchmark Dataset: MalwareBench}} \label{malwarebench}
Recent studies~\cite{ohm2020backstabber,duan2020towards,guo2023empirical,datadog_pypi} constructed malicious package datasets. However, these studies did not incorporate neutral package data into their labeled datasets. Neutral packages are needed to improve the accuracy of detection techniques. To address the lack of labeled neutral packages, Zahan et al. constructed MalwareBench~\cite{zahan2024malwarebench}, a labeled benchmark dataset of malicious and neutral npm packages. Zahan et al. refer to a package as \textit{neutral} when it has no discovered malicious code~\cite{zahan2024malwarebench}. %
The ground-truth labels for the benchmark were established using a hybrid approach: researchers manually reviewed and confirmed all malicious cases flagged by automated tools, while neutral cases were initially identified through automation and further validated via manual analysis of randomly selected samples. MalwareBench contains 3,523 malicious and 10,691 neutral npm packages. The malicious category comprises real-world malicious packages, while the neutral category includes popular, new, and randomly annotated real-world npm packages. %

\subsection{\textbf{Large Language Model Selection}} \label{LLmodel}
LLMs based on the GPT architecture have demonstrated superior performance in inference tasks compared to BERT-style models ~\cite{qin2023chatgpt,zhong2023can,yang2023harnessing,liu2023summary}. Inference tasks involve using a model's trained knowledge with new data to generate conclusions. While other GPT-style language models exist, %
ChatGPT has shown noteworthy capabilities in zero-shot learning through effective prompting and in-context learning~\cite{yang2023harnessing}. In this study, we employed the \texttt{gpt-3.5-turbo-1106} (GPT-3) and \texttt{gpt-4-1106-preview} (GPT-4) models of ChatGPT to evaluate LLMs' performance in detecting malicious packages using zero-shot learning techniques. 

\subsection{\textbf{Cost Allocation}} \label{LLM_Cost}

Due to the cost and time associated with using GPT models, we first allocated a budget for this project. %
The budget will inform the scope of our analysis regarding the number of files or packages that can be assessed. We set up a budget of \$3000 because increasing the budget would extend the model's computation time and the time needed for authors to manually review the model's generated response. %
We conducted a preliminary analysis of 1000 Javascript files using both GPT-3 and GPT-4 following the workflow mentioned in Section \ref{RQ1}. We found that analyzing 1000 Javascript files with GPT-3 incurred a cost of \$10.5, while the same analysis with GPT-4 cost \$130. Armed with this cost information, we deduced that with a budget of \$3000, we could analyze around 20,000 files.

\subsection{\textbf{Package Selection}} \label{package_selection}
In this work, we evaluated all files within a package rather than only analyzing files containing malicious code. The approach was taken for two reasons: to address the lack of file level granularity in MalwareBench~\cite{zahan2024malwarebench} and to understand false-positive and false-negative rates in our baseline comparison comprehensively. The dataset comprised 14,214 npm packages with 1,048,576 individual files. Packages varied in size, with some containing up to 60,000 files and occupying 600 MB. Analyzing exceptionally large packages could exceed our budget and may not provide sufficient diversity of malicious or neutral package types. Therefore, we excluded large packages to stay within budget and ensure diversity. Using stratified inverse random sampling~\cite{latpate2020inverse}, we selected a representative sample of malicious and neutral categories, favoring packages below the 75th percentile. The 75th quartile encompasses packages with a maximum of 25 files or a package size of 175 KB.
After applying the sampling method and adhering to budget constraints (section \ref{LLM_Cost}), our final dataset comprised \malwareversion malicious versions (\malwarepackage packages) and \neutralversion neutral versions (\neutralpackage packages), containing \totalfile (\totaluniquefile unique files) files. %

\begin{table*}[!htb]
\centering
\caption{\textbf{System Role Prompt of Initial Report}}\label{tab:initial_report_prompt}
\begin{tabular}{|p{80pt}|p{408pt}|} 
\hline
\textbf{Role: System} & \textbf{Chain of Thought Prompting} \\
\hline
Task (D1) &  As SecureGPT, a JavaScript cybersecurity analyst, your task is to review open-source dependencies in client and server-side JavaScript code for potentially malicious behavior or sabotage.
This code review is specifically for JavaScript libraries that are part of larger projects and published on public package managers such as npm. %
Review this code for supply chain security attacks, malicious behavior, and other security risks. Keep in mind the following: \\
& Analyze code for JavaScript security issues such as code injection, data leakage, insecure use of environment variables, unsafe SQL, and random number generation. Do NOT alert on minified code that is a result of standard minification processes using tools like UglifyJS or Terser. Third-party library usage is not, by itself, suspicious behavior.\\
\hline
Guidelines (D2) &  Spot anomalies: hard-coded credentials, backdoors, unusual behaviors, or malicious code. Watch out for malicious privacy violations, credential theft, and information leaks. Note observations about the code. Evaluate the provided file only. Indicate low confidence if more info is needed. Avoid false positives and unnecessary warnings. Keep signals and reports succinct and clear. Consider user intent and the threat model when reasoning about signals. Focus on suspicious parts of the code. \\
\hline
Malware (D3)& In the context of an npm package, malware refers to any code intentionally included in the package that is designed to harm, disrupt, or perform unauthorized actions on the system where the package is installed.\\
\hline
Malware Example (D4): & Sending System Data Over the Network, Connecting to suspicious domains, Damaging system files, Mining cryptocurrency without consent, Reverse shells, Data theft (clipboard, env vars, etc), Hidden backdoors. \\
\hline
Security risks (D5): & Hardcoded credentials, Security mistakes, SQL injection, DO NOT speculate about vulnerabilities outside this module. \\
\hline
Obfuscated code (D6): & Uncommon language features, Unnecessary dynamic execution, Misleading variables, DO NOT report minified code as obfuscated.\\
    \hline   
Malware score (D7): & 0: No malicious intent,\\& 0 - 0.25: Low possibility of malicious intent, \\&0.25 - 0.5: Possibly malicious behavior,\\& 0.5 - 0.75: Likely malicious behavior, e.g., tracking scripts,\\& 0.75 - 1: High probability of malicious behavior; do not use. \\
\hline
Security risk score (D8): & 0 - 0.25: No significant threat; we can safely ignore, \\&0.25 - 0.5: Security warning, no immediate danger,\\& 0.5 - 0.75: Security alert should be reviewed,\\& 0.75 - 1: Extremely dangerous, package should not be used. \\
\hline
Confidence Score (D9) & Rate your confidence in your conclusion about whether the code is obfuscated, whether it contains malware, and the overall security risk on a scale from 0 to 1.\\
\hline
Code Review (D10): & Please consider both the content of the code as well as the structure and format when determining the risks. Your analysis should include the following steps:\\
 & \textbf{Identify sources}: These are places where the code reads input or data.\\
 & \textbf{Identify sinks}: These are places where untrusted data can lead to potential security vulnerabilities.\\
 & \textbf{Identify flows}: These are source-to-sink paths.\\
& \textbf{Identify anomalies}: These are places where there is unusual code, hardcoded secrets, etc. \\
& \textbf{Conclusion}: Finally, form the conclusion of the code, provide a succinct explanation of your reasoning.\\
\hline
JSON Response (D11)& **Only respond in this format:** \\
&[\\&``\textbf{purpose}":``Purpose of this source code",\\
&``\textbf{sources}":``Places where code reads input or data",\\
&``\textbf{sinks}":``Places where untrusted data can lead to potential data leak or effect",\\
&``\textbf{flows}":``Source-to-sink paths",``\textbf{anomalies}":``Places where code does anything unusual",\\&``\textbf{analysis}":``Step-by-step analysis of the entire code fragment.",\\&``\textbf{conclusion}":``Conclusions and short summary of your findings",\\&``\textbf{confidence}": 0-1,\\&``\textbf{obfuscated}": 0-1,\\&``\textbf{malware}": 0-1,\\&``\textbf{securityRisk}": 0-1\\
&]\\
&ONLY RESPOND IN JSON. No non-JSON text responses. Work through this step-by-step to ensure accuracy in the evaluation.\\
  \hline
\end{tabular}
\end{table*}

\begin{table*}[!htb]
\caption{\textbf{System Role Prompt of Critical Reports}}\label{tab:Critic_Analysis} 
\centering
\begin{tabular}{|p{80pt}|p{408pt}|} \hline
\textbf{Role:System} & \textbf{Chain of Thought Prompting} \\\hline
Task & Your role is a security researcher who will review $X$ reports of a supply chain security incident in an open source dependency. Here's your step-by-step guide: \\
& - Confirm if the reported issues are present in the code. If not, make a note. \\
& - Identify and record any errors, flaws or mistakes in the report's logic, reasoning or summary.\\
& - Review the scores given to each issue. Suggest any changes if they seem unreasonable.\\
& - Justify any risk score higher than 0.5, as it's considered serious.\\
& - Pay close attention to the malware score. If a package is doing something malicious, it must be reflected in this score.\\
& - Challenge any claims about potential vulnerabilities that are based on assumptions.\\
& - Make sure the scores are consistent with the report.\\
\hline
Note & Be critical. After reading all reports, give your own estimate for what you think the malware, obfuscated, and risk scores should be. Let's carefully work this out step-by-step to be sure we have all the right answers.\\
\hline
\end{tabular}
\end{table*}

\begin{table*}[!htb]
\caption{\textbf{System Role Prompt of Final Report}}\label{tab:Summary_Report_Prompt} 
\centering
\begin{tabular}{|p{80pt}|p{408pt}|} \hline
\textbf{Role:System} & \textbf{Chain of Thought Prompting} \\\hline
Task & As an experienced JavaScript cybersecurity analyst, review $X$ software supply chain security reports of an open source dependency. Select the best report and improve it. If all are unsatisfactory, create a new, improved summary. \\
& System Role Prompt of Initial Reports from Table \ref{tab:initial_report_prompt}.\\
\hline
Note & In your final report: \\&- Omit irrelevant signals from your report. \\&- If you are certain code is really malicious set the malware score to 1. \\&-ENSURE THAT YOUR FINAL SCORE IS CONSISTENT WITH THE REPORT. \\
& **ONLY RESPOND IN JSON. PLEASE DO NOT USE ANY TEXT BESIDES THE JSON REQUIRED FOR THE FINAL FORMATTED REPORT.**  Let's work this out in a step-by-step way to be sure we get the correct result.\\
\hline
\end{tabular}
\end{table*}

\section{\textbf{\SocketAIScanner}} \label{SocketAIScanner}
The section discusses the multi-stage decision-maker workflow of \SocketAIScannernogap. \socket has been using \SocketAIScanner to detect malicious packages in production for npm, PyPI, Go, and Maven ecosystems. On average, \SocketAIScanner is used to evaluate 17 million package versions quarterly in production. As of 2024, three authors are actively developing and maintaining the workflow. In Section \ref{Iterative_Self_Refinement}, and \ref{prompt}, we discuss our design decisions and the prompting techniques, and section \ref{model_workflow} discusses our proposed \SocketAIScanner workflow.

\subsection{\textbf{Leveraging LLM-Based Techniques}} \label{Iterative_Self_Refinement} \label{CoT}
Since \SocketAIScanner is used in production to review different ecosystems, we designed our domain-specific prompt to be model-agnostic and adaptable to open and closed-source models where the prompt techniques are fundamental to any LLM. Our prompting techniques were inspired by advanced LLM-based techniques by~\cite{madaan2023self, zheng2023judging,kojima2205large,wei2022chain,ye2023prompt}, while also prioritizing a domain-targeted prompt using the author's domain knowledge.  Prior studies~\cite{ahmed2024automatic,white2024chatgpt, white2023prompt} demonstrate that augmenting the model with domain-specific information improves the model's performance.

The \SocketAIScanner workflow employs an iterative self-refinement~\cite{madaan2023self} approach, systematically leveraging LLM's responses in a sequential analytical workflow. Iterative self-refinement is a process that involves creating an initial draft response generated by the model and subsequently refining the response based on the model's self-provided feedback~\cite{madaan2023self, zhang2023prefer}. Madaan et al.~\cite{madaan2023self} showed that iterative self-refinement prompting had superior results compared to one-step response generation and improved performance (20\%). Then, we also leveraged the LLM-as-a-Judge approach~\cite{zheng2023judging} to evaluate and score the outputs against specific criteria. Self-refinement is a loop of generation and improvement, while LLM-as-a-Judge is better for scoring or verifying outputs against criteria. We used the Zero-Shot-CoT~\cite{kojima2205large,wei2022chain,ye2023prompt} technique along with Role-Play prompting~\cite{kong2023better} to enhance model reasoning in evaluating npm packages. Prior studies~\cite{kojima2205large, kong2023better, ye2023prompt} showed that Zero-Shot-CoT and Role-Play prompting enhance model performance across multiple domains. Subsequently, in Role-Play prompting, the role provides context about the model’s identity and background, which serves as an implicit CoT trigger, thereby improving the quality of reasoning.

\begin{figure*}[!htb] 
  \centering
    \includegraphics[width=\linewidth]{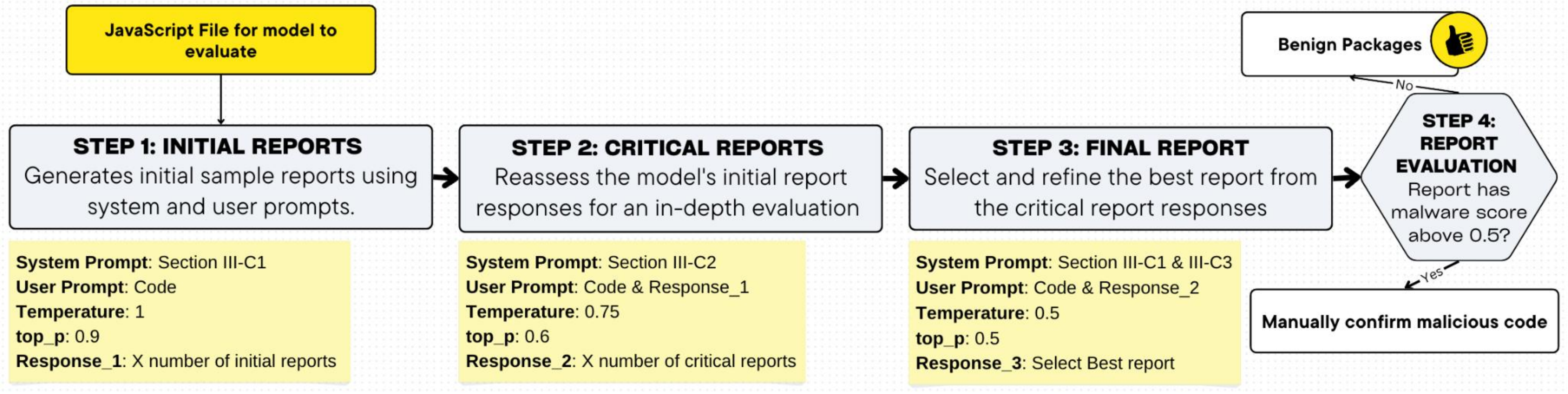}
    \caption{\SocketAIScanner workflow}
    \label{fig:prompt_workflow}

\end{figure*}

\subsection{\textbf{Prompt Design}} \label{prompt}
We designed a two-fold prompt structure: the \texttt{system role} prompt serves as the instructions to set the context for how the model should behave. We assign different \texttt{system roles} for LLMs to follow in multiple steps (Section \ref{model_workflow}) such as \textit{SecureGPT, a JavaScript cybersecurity analyst} and \textit{security researcher}, to narrow the focus to JavaScript’s particular security issues. The \texttt{user role} prompt comprises the input content (npm packages files) we want the models to review. For each file, we include the code as a user prompt, following the prompt\textemdash~\textit{Analyze the above code for malicious behavior. Remember to respond in the required JSON format. Consider ALL of the code carefully. Check the beginning, middle, and end of the code. Work step-by-step to get the right answer}.  Table \ref{tab:initial_report_prompt}, Table \ref{tab:Critic_Analysis} and Table \ref{tab:Summary_Report_Prompt} contain the prompts used in this study. 

\textbf{Task \& Guidelines. }
We first augment the prompt by defining the task of reviewing JavaScript code for supply chain attacks, malicious behavior, and other security risks.  Then, we provide guidelines to spot anomalies, such as hardcoded credentials, backdoors, and unusual behaviors, while noting code observations and avoiding unnecessary warnings.

\textbf{Malware, Security Risk, and Obfuscated Code.}
We provide definitions and examples for malware, security risks, and obfuscated code. The targeted information helps the model recognize and differentiate between malware and other security vulnerabilities. Additionally, in malware detection, one of the primary challenges is prioritizing and differentiation between malware cases and security vulnerabilities. Motivated by LLM-as-a-Judge approach~\cite{zheng2023judging}, we address this challenge by incorporating scoring metrics— \texttt{malware intent}, \texttt{security risk}, and \texttt{confidence} — to standardize the evaluation output, enabling the model to quantify the severity of findings and assisting human reviewers in prioritizing potentially malicious files. Since attackers commonly use \texttt{obfuscation} to avoid detection, it is included to identify such cases.
The confidence score also aids in evaluating the reliability of the model's findings, while malware, security risk, and obfuscation are included to understand the nature of the code. Each scoring is defined in a range of 0-1, which we divide into different groups. Table \ref{tab:initial_report_prompt} (D7-D9) are examples of scoring prompts.

\textbf{Requirement: Code Review \& JSON Response.}
Motivated by \cite{ahmed2024automatic,white2024chatgpt, white2023prompt}, we included code-review prompts (Table \ref{tab:initial_report_prompt}-D10) that guide the model in step-by-step thinking and generating conclusions by detecting key identifiers of code such as \texttt{sources, sinks, flows, anomalies}. Additionally, large-scale analysis in production requires machine-readable reports. Therefore, we improved the prompting technique by instructing the model to conduct a sequential analysis (\texttt{purpose-sources-sinks-flows-analysis-conclusion}) and to generate the output report via a JSON format~(Table \ref{tab:initial_report_prompt}-D11). Besides the report being machine-readable, following such templates forces the model to rethink its evaluation and generate analysis reports in the specified format. %

\textbf{Token Limit}: GPT-3 has a maximum of 16K tokens, and GPT-4 has a limit of 128K tokens. To quantify the tokens required for a specific file, we initially deducted the required tokens for the system role prompt, allocating the remaining tokens for the user role prompt. %

\subsection{\textbf{Proposed Workflow}} \label{model_workflow}
Our \SocketAIScanner workflow comprises four steps (Figure \ref{fig:prompt_workflow}): code as user input along with a system role prompt (Step 1); refining the model's responses through iterative self-refinemnet in the next two steps (Steps 2 and 3); and finally, adding humans in the loop to evaluate the model-generated report~(Step 4). %

\subsubsection{\textbf{Step 1: Initial Reports}} \label{Initial_Reports}
We provide the file to evaluate as user role prompts alongside predefined system role prompts from Table \ref{tab:initial_report_prompt} to generate initial response reports. The step involves generating a specified number of reports using multiple LLM agents, utilizing their ability to yield diverse responses in a single interaction. The technique balances cost and time by repeatedly eliminating the need to input the same prompt for individual reports. In this study, we prompted GPT-3 to generate 5 reports per file. We used GPT-4 to generate 3 reports per file due to its higher accuracy and cost. The model is tuned with a temperature parameter of 1 and a $top_p$ value of 0.9 to explore a wide range of ideas. %
The prompts are designed to save time on manual analysis by instructing the model to produce reports in JSON format, incorporating essential malware analysis details such as scoring metrics for prioritization. The output of this step consists of 5 reports generated by GPT-3 and 3 reports generated by GPT-4, which we then feed into Step 2 as user role prompts (Figure \ref{fig:prompt_workflow}).  %

\begin{table*}[!htb]
\centering
\begin{tabular}{|p{0.1\linewidth}||p{0.85\linewidth}|}
\hline
\textbf{} & \textbf{Shell Scripts for GPT-4 Evaluation} \\
\hline
Corporate-delegate-packages & 
\begin{minipage}[t]{\linewidth}

\begin{lstlisting}[basicstyle=\ttfamily\small, frame=none]
#!/bin/bash
curl -H "Hostname: $(hostname | base64)" -H "Whoami: $(whoami | base64)" -H "Pwd: $(pwd | base64)" -d $(ls -la | base64) https://eo1z2prk4as56mb.m.pipedream.net
\end{lstlisting}
\end{minipage} \\
\hline
\hline
\hline
\textbf{} & \textbf{GPT-4 Generated Final Report} \\
\hline
Purpose (R1) & The purpose of this source code is to covertly exfiltrate sensitive system information to a remote server, which is indicative of a malicious script. \\
\hline
Sources (R2)& hostname command, whoami command, pwd command, ls -la command \\
\hline
Sinks (R3) & curl command \\
\hline
Flows (R4)& \texttt{hostname} command $\rightarrow$ Hostname HTTP header, 
\texttt{whoami} command $\rightarrow$ Whoami HTTP header,
\texttt{pwd} command $\rightarrow$ Pwd HTTP header,
\texttt{ls -la} command $\rightarrow$ HTTP POST data\\
\hline
Anomalies (R5) & Base64 encoding of system information, Incomplete URL in the curl command, Sending system information via HTTP \& POST \\
\hline
Analysis (R6)& \textbf{$\rightarrow$} Uses the hostname, whoami, and pwd commands to collect the system's hostname, current user, and current working directory., \textbf{$\rightarrow$} ls -la command is used to list all files and directories in the current working directory, which may include sensitive information, \textbf{$\rightarrow$} All collected data is base64 encoded (a weak form of obfuscation) to avoid simple detection mechanisms., \textbf{$\rightarrow$} The encoded data is sent to a remote server using the curl command with custom HTTP headers and POST data, which indicates data exfiltration.\\
\hline
Conclusion (R7)& The code exhibits behavior consistent with a malicious script designed for data exfiltration. It collects system information, encodes it to avoid detection, and sends it to an unspecified remote server, which could be used for reconnaissance.\\
\hline
Score (R8)& \textbf{Confidence}: 0.9, \textbf{Malware}: 1, \textbf{SecurityRisk}: 1, \textbf{Obfuscated}: 0.2\\
\hline
\end{tabular}
\caption{Example of the user input file for model to analyze and GPT4 generated final report}
\label{tab:CHATGPT_response}

\end{table*}

\subsubsection{\textbf{Step 2: Critical Reports}} \label{critical_reports}

The critical reports step is designed to reassess the model's initial report responses to perform an in-depth evaluation. As input, the system prompt (Table \ref{tab:Critic_Analysis}) instructs the model, acting as a security researcher, to review and evaluate the initial findings thoroughly. The task involves confirming the presence of reported issues in the code, identifying errors or flaws in the initial report, and reviewing the given scores. The model is asked to justify any risk score above 0.5, ensure the score accurately reflects malicious behavior, challenge assumptions about vulnerabilities, and maintain consistency in scoring. For the user role prompt, we input the model-generated responses from Step 1~(section \ref{Initial_Reports}) and the file to be evaluated. The model generates a predetermined number of output reports (5 for GPT-3 and 3 for GPT-4). The model is tuned to a temperature parameter of 0.75 and a $top_p$ value of 0.6 to constrain the idea generation.

\subsubsection{\textbf{Step 3: Final Report}} \label{final_reports}

The final report prompt~(Table \ref{tab:Summary_Report_Prompt})  is designed to select and refine the best report from step 2. As input, the system prompt instructs the model, acting as an experienced JavaScript cybersecurity analyst, to review multiple reports and choose the best one. The task involves eliminating irrelevant signals by reviewing reports from step 2 and setting the malware score to 1 if the code is malicious to emphasize consistency between the final score and the report content. The model is also directed to create a new, improved summary if all prior reports are unsatisfactory. The step also involves sequentially feeding the model system role prompts from Step 1~(Table \ref{tab:initial_report_prompt}) to remind the objectives and ensure a thorough code reevaluation. %
Overall, the model uses system role prompts (Table \ref{tab:Summary_Report_Prompt} and Table \ref{tab:initial_report_prompt}), the input content (npm packages files), and reports from step 2 to output a conclusive summary report, operating at a lower temperature of 0.5 and a $top_p$ value of 0.5 to enhance focus and precision. %

\subsubsection{\textbf{Step 4: Report Evaluation}} \label{Manual_Evaluation} In this step, a human evaluates the final report generated in step 3. Table~\ref{tab:CHATGPT_response} contains the shell scripts of real-world malicious packages ``\texttt{corporate-delegate-packages}'' and the associated GPT-4-generated final report. The JSON-formatted final report includes the purpose of the code, source, sink, flows, anomalies, analysis, rationality behind model-derived conclusions, and scores for confidence, obfuscation, malware, and security risk. Reviewers can prioritize evaluations based on the malware score. Based on our scoring rules, a package is prioritized for human evaluation if any file has a malware score exceeding 0.5 (Section \ref{RQ2_ChatGPT}). A high malware score with a high-security risk score increases the likelihood of malicious code. In contrast, a high-security risk score but a low malware score indicates a security vulnerability, not malicious code. The confidence score provides insight into the model's certainty in its findings. An obfuscated score is helpful to understand whether the attacker used obfuscation to hide the code intent. Reviewers can review the purpose and conclusion parts (R1 \& R7 in Table ~\ref{tab:CHATGPT_response}) to comprehend the code's intent. If further verification is needed, they can analyze the code review metrics (R2-R6 in Table \ref{tab:CHATGPT_response}).   %

\section{RQ1: Baseline Comparison} \label{RQ1}

In this section, we discuss RQ1: \textbf{\RQOne} To that end, we adopt a traditional static analysis that is used to extract static features from the code and GPT-3 and GPT-4 for LLM analysis. 
Section \ref{RQ1_methodology} covers our methodology, and Section \ref{RQ1_result} covers our result.
\subsection{\textbf{RQ1: Methodology}} \label{RQ1_methodology}

\subsubsection{\textbf{CodeQL Static Analysis}} 
\label{RQ2_SAST_feature_extraction} 
For our baseline comparison, we selected the static analyzer over any other existing technique to reduce the production cost of pre-screening. We chose not to adopt other existing static analyzers focused on malicious detection due to several factors: 1) a high false positive rate~\cite{pfretzschner2017identification, vcarnogursky2019attacks}, reliance on combined detection techniques such as static, dynamic, and machine learning~\cite{duan2020towards}, whereas we used static rules only to minimize overhead, and 3) the tools were tailored to detect malicious commits in VCS~\cite{gonzalez2021anomalicious} or targeted languages like Python~\cite{milje2022detecting}  and Java~\cite{ladisa2022towards}, rather than our focus on npm.

Froh et al., in a recent 2023 study~\cite{froh2023differential}, introduced a differential static analyzer leveraging custom CodeQL queries to identify JavaScript malicious code. For our baseline comparison, we utilized CodeQL queries~(available in Github~\cite{diff-codeql}) developed by Froh et al. ~\cite{froh2023differential}. We utilized CodeQL over other existing static tools because 1) CodeQL is a commercially mature, state-of-the-art static analyzer, open source, and more developed than other static analyzers proposed in research for malicious package detection~\cite{ohm2023sok}; 2) the CodeQL queries used by Froh et al. [7]  were explicitly designed for npm malicious packages, aligning with our dataset selection.

We used 39 out of 40 queries developed by Froh et al., excluding the query, to detect the introduction of new dependencies.  Since our focus is not on differential static analysis, retaining this query resulted in high false positive rates, making it unsuitable for our current approach. The 39 remaining queries include \textit{code injection}- involving the use of functions like eval for executing dynamically generated code; \textit{sensitive data exposure}- focusing on unauthorized access to sensitive user and system files; \textit{network issues}- identifying unauthorized data flows potentially indicating exfiltration or backdoor attempts; \textit{file system access}- monitoring access to files that may indicate security breaches; \textit{obfuscation and encoding}- detecting code obfuscation techniques used to hide malicious intent; and \textit{miscellaneous}- covering other concerns such as unsafe handling of domain names. %

\subsubsection{\textbf{ChatGPT Models}} \label{RQ2_ChatGPT}

We leverage the \SocketAIScanner workflow discussed in Section \ref{SocketAIScanner} to review \totaluniquefile unique files using GPT-3 and GPT-4. We ran both models on each file and then collected the final reports discussed in Section \ref{final_reports}. 
According to our evaluation criteria (Section \ref{Manual_Evaluation}), a malware score of 0.5 marks the threshold between 'possible malware behavior' (0.25-0.5) and 'likely malicious behavior' (0.5-0.75). We aimed to assess the accuracy of LLMs in generating these scores, considering a package malicious if any file within it scored above 0.5.

\subsection{\textbf{RQ1: Results}} \label{RQ1_result}
Here, we discuss our baseline comparison and show confusion metrics and model performance in Table~\ref{tab:model_performance}. %

\begin{table}[!htb]
\centering
\caption{Performance Evaluation using \uniquepackage packages}
\begin{tabular}{|p{88pt} || p{25pt}| p{25pt}| p{22pt}| p{22pt}|}
\hline
Model  & TP & TN & FP  & FN \\
\hline\hline
CodeQL  & 2117 & 2254 & 684  & 60 \\
\hline
GPT-3  & 2128 & 2740 & 195  & 52  \\
\hline
GPT-4 & 2089 & 2932 & 3  & 90  \\
\hline
\end{tabular}
\begin{tabular}{|p{88pt} || p{35pt}| p{35pt}| p{36pt}|}
\hline
  & \textbf{Precision} & \textbf{Recall} & \textbf{F1} \\
\hline\hline
CodeQL & 0.75 & 0.97 & 0.85 \\
\hline
GPT-3 & 0.91 & 0.97 & 0.94 \\
\hline
GPT-4 & 0.99 & 0.95 & 0.97 \\
\hline
\end{tabular}
\label{tab:model_performance}
\end{table}

\subsubsection{\textbf{CodeQL's Performance}} \label{SAST_performence}
After executing CodeQL on the dataset, we found \sastpackage packages with at least one file with suspicious features. CodeQL queries had high coverage in identifying features in malicious packages, with a recall rate of 0.97. A high recall is expected since the queries constructed by Froh et al.~\cite{froh2023differential} utilized the same past malicious package dataset that we used in our study. However, the model correctly identified 97\% of malicious packages but at the cost of potentially incorrectly labeling many non-malicious instances as malicious. The model had successfully classified 2,117 malicious instances (true positives) and 2,254 instances as neutral (true negatives) out of \uniquepackage unique packages. The model has a false positive of 684 neutral packages that demanded other techniques or manual analysis to confirm the classification. The precision rate is 0.75, and the F1 score of 0.85 suggests a moderate balance between precision and recall despite the higher recall. %

\subsubsection{\textbf{GPT Model's Performance}} \label {GPT_performence}

The GPT-3 model demonstrates an improvement over static analysis, with a precision of 0.91 and an F1 score of 0.94, identifying 2,128 malicious and 2,740 neutral cases accurately (Table~\ref{tab:model_performance}). GPT-3 exhibits a lower false positive rate (195) than CodeQL, highlighting the model's reliability. A recall rate of 0.97 indicates the model captures 97\% of malicious instances. The GPT-4 model further advances performance, achieving the highest precision (0.99) and an F1 score of 0.97, correctly classifying 2,089 malicious cases and 2,932 neutral cases, with significantly fewer false positives (3) and a recall rate of 0.95. This comparison highlights GPT-4's superior ability to balance precision and recall, making it an effective model for distinguishing malicious and neutral packages. %

\textbf{Summary of RQ1} Both GPT-3 and GPT-4 models demonstrate superior performance in precision and F1 score for malicious code detection. GPT-4 has the highest false negative rate. Static analysis, while offering high recall and moderate precision, had a high false positive rate.%

\section{RQ2: Effectiveness of \SocketAIScanner} \label{RQ2}

In this section, we discuss RQ2: \textbf{\RQTwo}

\subsection {\textbf{RQ2: Methodology}} \label{RQ2_method}

An LLM workflow can be expensive if practitioners want to review at scale (in millions of \$ for GPT-4). Additionally, OpenAI has an API call rate limits~\cite {API_rate}. Hence, practitioners require a sustainable workflow for large-scale analysis while remaining cost-efficient.  Our study investigates the effectiveness of static analyzers as a pre-screening step before our LLM workflow. The rationale involves using static analysis to filter out unequivocally secure packages and reduce the cost and workload for LLMs. %

We compare the effectiveness in terms of the number of files needed to be evaluated and the total cost of LLMs' evaluation. We designed the experiment in which we first scanned all files (a total of \totaluniquefile files) using both GPT models and calculated the associated costs. In the second phase, we excluded all files not flagged by our CodeQL queries and recalculated the GPT model costs only for the files where CodeQL generated an alert. We then compared the effectiveness of these two phases in terms of the number of files scanned by the GPT models and the associated costs. 

\begin{figure}[!htb] 
  \centering
    \includegraphics[width=\linewidth]{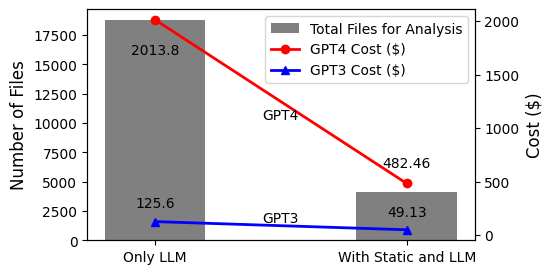}
    \caption{Effectiveness of using static analysis as pre-screener}
    \label{fig: SAST_LLM}
\end{figure}

Note that we did not conduct a detailed time analysis, as the response times for both GPT models were similar ( 24 days per GPT model and approximately 190 hours of qualitative study per independent researcher) and often influenced by OpenAI's performance, including the need for re-runs due to technical issues. CodeQL took one day to run all the queries. Since we did not thoroughly evaluate false positives and negative cases of the static analyzer used in the prior study~\cite{froh2023differential}, comparing CodeQL with GPT run time did not provide a balanced assessment. Figure \ref{fig: SAST_LLM} shows our comparison of file and cost analysis.

\subsection{\textbf{RQ2: Result}} \label{RQ2_result}
\subsubsection{\textbf{GPT Model's file Analysis}} 
As a pre-screening step, CodeQL is used to identify files that should be flagged for further LLM analysis. Out of \totaluniquefile unique files, \sastfile files were flagged by CodeQL. 

\subsubsection{\textbf{GPT Model's Cost Analysis}} A cost comparison of the GPT-3 and GPT-4 models demonstrates substantial differences in expense. To scan a total of \totaluniquefile unique files, the GPT-3 model incurred a cost of \$125.65. In contrast, the GPT-4 model, with its advanced capabilities and presumably superior performance \cite{openai}, cost \$2013.84 for the same number of files. Scanning the \sastfile files flagged by CodeQL cost \$49.13 with GPT-3 and \$482.46 with GPT-4. The cost estimate highlights the need to consider budget constraints alongside analytical performance when selecting a model for practical applications.

\textbf{Summary of RQ2: }Pre-screening files with a static analyzer reduces the number of files needing analysis from 18,754 to 4,146, demonstrating a 77.9\% reduction. This pre-filtering step also substantially decreases costs, with GPT-3 expenses dropping from \$125.65 to \$49.13 (60.9\%) and GPT-4 costs reducing from \$2013.84 to \$482.46 (76.1\%). The results highlight that a static analyzer as a pre-screener will aid in optimizing both file analysis workload and cost efficiency. %

\section{RQ3: Manual Evaluation of LLMs analysis} \label{RQ3}
In this section, we answer our RQ3: \textbf{\RQThree} For our study, we needed the ground truth at the file level. However, MalwareBench contains granularity at the package level. Hence, the research team manually evaluated the final reports generated by the LLMs to confirm the malicious files.  

\subsection{\textbf{RQ3: Methodology}}\label{manual_analysis}
To answer RQ3, we use a sorting approach to facilitate manual evaluation, where two researchers independently assess the model's generated report (Table ~\ref{tab:CHATGPT_response}). Since we had \totaluniquefile unique files, manually evaluating each file would be time-consuming.  Therefore, we screened files based on the malware score (Table \ref{tab:CHATGPT_response}-R8). We discuss our approach below- %

\textbf{Malicious Category}: We manually evaluated 2,301 unique files (2,226 from GPT-3 and 1,025 from GPT-4) that had a malware score above 0.5. %
We conducted an additional manual analysis of 123 malicious packages to check for false negative cases that are missed by LLMs. Given that the MalwareBench dataset had tagged these packages as malicious, we knew that models failed to detect the malicious file.
\textbf{Neutral category}: We manually evaluated 107 neutral package files. Since prior research~\cite{zahan2024malwarebench} classified these packages as neutral, we manually assessed the model analysis to determine whether they were false positive cases missed by LLM. %

\subsection{\textbf{RQ3: Results}} \label{Rq3_result}
We first discuss our true positive or malicious cases in Section \ref{TP}. Section \ref{FP} covers the cases of false positives, where the models incorrectly identified non-malicious cases as malicious. Section \ref{FN} covers false negatives, where the models failed to identify an actual malicious case.
\subsubsection{\textbf{True Positive (TP)}} \label{TP}
In our analysis, the overlapping set of malicious packages detected by both models comprised 2,057 out of a total of \malwarepackage malicious packages. GPT-3 and GPT-4 independently detected 2,128 and 2,089 packages, respectively. Our manual evaluation by two researchers further revealed a significant concentration of malicious packages, primarily targeting data theft or exfiltration (736 cases) to harvest sensitive user information. The analysis also indicated 70 instances of reverse shells, facilitating unauthorized remote access, and 126 instances of hidden backdoors embedded within packages for system entry. Additionally, the models detected 391 cases of malicious connecting to suspicious domains, including \texttt{Pastebin} and domains controlled by attackers. Typosquatting was identified in 19 instances, exploiting user errors by registering misleading domain names to distribute malware. Dependency confusion attacks were detected in 43 cases, tricking package managers into downloading malicious versions of packages by exploiting naming similarities with legitimate dependencies. 

We summarize our attack type in Table~\ref{tab:malware_summary}. Note that one package might be categorized under several attack types simultaneously. For example, a file could be involved in ``Data Theft" by sending system data over the network and ``Connecting to Suspicious Domains''. 

\begin{table}[ht]
\centering
\captionsetup[table]{skip=2pt}
\caption{Summary of Detected Malicious Packages}
\label{tab:malware_summary}

\begin{tabular}{|p{140pt} || p{70pt}|}
\hline
\textbf{Attack Type}              & \textbf{Package Count} \\ \hline
Data Theft or Exfiltration        & 736            \\ \hline
Hidden Backdoors                  & 126        \\ \hline
Connecting to Suspicious Domains  & 391            \\ \hline
Reverse Shells                    & 70           \\ \hline
Execution of Arbitrary Code       & 243           \\ \hline
Dependency Confusion              & 43            \\ \hline
Typosquatting                     & 19             \\ \hline
\end{tabular}
\end{table}

 Here we give three examples of GPT-4 generated conclusions for three malicious files: \textbf{EX1}: \textit{The code is likely malicious, as it appears to exfiltrate sensitive system and user information through DNS queries and HTTPS requests to a domain that resembles known Interactsh domains used for malicious purposes.} \textbf{EX2}: \textit{The `preinstall' script could be a vector for malicious activity...The name `escape-htlm' could be a typo or a potential attempt at typosquatting...}, \textbf{EX3}: \textit{The package.json exhibits several red flags, including a likely typosquatting attempt and an `preinstall' script that could execute arbitrary code.}%

We observed GPT-3 assigned malware scores above 0.5 to a significantly higher number of files (2,226 files) compared to GPT-4 (1,025 files), resulting in 1,191 additional files requiring manual review. Among these additional files,  94\% of the files were package.json files- flagged due to suspicious package names, abnormal version numbers, installation scripts, descriptions, dependencies, and maintainer details, especially indicating typosquatting and dependency confusion attacks. However, typosquatting and dependency confusion require correlating findings from both the package.json file and the source code to determine the attack's type. In such cases, GPT-4 assigned more moderate malware scores (between 0.25 and 0.5) to such files and detected the actual issues in the source code, indicating a more accurate assessment. This mismatch in scoring and reasoning highlights the GPT-3's inference based on incomplete evidence. We tagged these packages as true positive because GPT-3 also correctly identified actual malicious files. However, high malware scores for non-malicious files indicate GPT-3 frequently hallucinates, and additional files require manual review. Further discussion on hallucinations can be found in Sections \ref{FP} and \ref{FN}.

\subsubsection{\textbf{False Positives~(FP)}} \label{FP} \label{FP}
GPT-3 identified 195 neutral packages as malicious packages exhibiting signs of potential error in the pattern recognition capabilities of GPT-3. Our manual analysis showed that \textbf{GPT-3 flagged} legitimate URL sanitizer code, use of post-installation scripts, dynamic code execution, and others as potential threats without adequately considering security vulnerabilities, code practices context, or legitimate use cases. We found 22 instances where GPT-3 incorrectly classified security vulnerabilities as malware, but GPT-4 accurately assigned a high-security risk but low malware score. Additionally, 173 packages were tagged as malicious by GPT-3 because of hallucination. Most of these input files were \textbf{empty or contained minimal content} of javascript files, \texttt{.md}, or \texttt{.json} files. For example, a \texttt{readme.md} file contain a text ``place holder'', GPT-3 concluded that\textemdash~\textit{The code is likely gathering system information and sending it to an external server. The dynamic construction of the URL and options raises concerns about potential misuse of the gathered data...}. On the contrary, GPT 4 concluded \textemdash~\textit{The provided content does not contain any JavaScript code to analyze; it's just placeholder text.} We suspect GPT-3 hallucinates because of a mismatch between expectations and insufficient information in the input. Such hallucination also highlights GPT-3's limitations over GPT-4 in iterative self-refinement prompting (Section ~\ref{Iterative_Self_Refinement}). Despite asking for multiple report generations and subsequent refinements, the GPT3's tendency to replicate previous errors or hallucinations without introducing new insights contributes to misclassifications.

GPT-4's had only 3 false positives. In one of the cases, the model flagged a \texttt{package.json} file as suspicious due to the inclusion of a new dependency version, \texttt{'eslint-plugin-import': 'npm:eslint-plugin-i@2.28.0-2'}, mistaking it for a potential dependency confusion attack. \texttt{eslint-plugin-i} is a fork of eslint-plugin-import to replace \texttt{tsconfig-paths}. Though GPT-4 showed attention in this instance, which is beneficial for security, it may still struggle with new information after its last training cutoff in April 2023. Then, GPT-4 incorrectly raised concerns about a module \texttt{http: 0.0.1-security} due to an unusual version number, which was likely a developer's error (trying to add a security holding npm packages as a dependency) rather than an intentional security threat.

\subsubsection{\textbf{False Negative (FN)}} \label{FN}
Our study had 142 false negative cases where both models did not identify the true malicious code. We have observed cases where the models correctly \textbf{detect security issues but assign a low malware score}. An exemplary case of GPT-4\textemdash~\textit{The code exhibits potentially malicious behavior by collecting sensitive user information and sending it to a remote location under certain conditions...} Despite these red flags, the model assigned a low-risk score of 0.4, suggesting a disconnect between the detection of individual risk factors and the aggregate assessment of malicious intent. 

Conversely, we have encountered cases where the models failed to identify actual security issues because \textbf{model overlooked the security issues or malicious code was incorrect, or the code appeared benign}. For example, models (both) concluded: \textit{The package.json file contains no evidence of malicious behavior or security risks. The configuration focuses on standard development, testing...}. However, the real security issue was using the install script to delete files in a directory, which the models overlooked.  Then, in one package, GPT-4 concluded that: \textit{ The package is suspicious due to the unconventional `preinstall' script invoking `burpcollaborator'....}. GPT-4 detected the issues but assigned a malware score of 0.3. The package malware intent was also unclear to the authors of this research because it only uses \textit{``preinstall":``burpcollaborator"}. Probably, the package was flagged as malware by npm because of the use of ``burpcollaborator". Then, in another case, the purpose of the malicious code was to sabotage one of the dependents from running successfully by preventing the download from completing~\cite{FN_ex}. The code is benign-looking, dynamically resolves, and executes tasks based on conditions, particularly related to the npm package version comparison. Therefore, GPT-4 assigned a malware score of 0 with a conclusion of \textit{The code does not appear malicious. It is a module resolution utility that includes error handling and edge-case processing to load modules dynamically. It does have potential security risks....}

We also observed cases with a high malware score, but the model \textbf{generated reports with the default JSON response} given in Table \ref{tab:initial_report_prompt}-D11, without including any new findings after evaluating the code. We received reports like \textit{``purpose": "Purpose of this source code"} or conclusions like \textit{"conclusion": "Conclusions and a short summary of your findings"}. %
Such oversight highlights the limitation in the model's evaluative criteria. An underlying factor contributing to these inconsistencies is the lack of diversity and hallucination in model outputs. Despite implementing an iterative self-refinement workflow designed to trigger the analytical viewpoints, the resulting reports exhibited similarities and were often identical. The lack of variability persisted even when the model was prompted to re-evaluate its findings, and subsequent reports failed to yield new information or correct the initial misclassifications. In the domain of LLMs, such limitations are known as ``mode collapse''~\cite{hamilton2024detecting}, a phenomenon where the model starts generating very similar or identical or noisy data, failing to capture the diversity of the original data~\cite{hamilton2024detecting,Shumailovcurse, briesch2023large}.  

\textbf{Summary of RQ3: }Both models exhibit superior performance in detecting true malicious files, with a low misclassification rate. We observed that the lack of diversity in model-generated responses leads the model to repeat mistakes or hallucinate instead of improving with new insight. Although GPT-3 hallucinates more frequently, the low number of incorrect classifications in terms of FP and FN also highlights the effectiveness of our LLM workflow. %

\section{Discussion }

In this section, we delve into the broader implications of using \SocketAIScanner and explore the challenges and threats to validity encountered in this study.

\subsection{\textbf{Implications of \SocketAIScanner and future work}: }Our contribution paves the way for new tool development and future research in AI-driven software development and software security domain. While prior studies use LLMs to detect vulnerabilities~\cite{purba2023software}, to our knowledge, ours is the first to propose an end-to-end LLM-based workflow for malicious code review. Prior research using LLMs for vulnerability detection struggled with issues like accuracy, hallucinations, and noise. We believe our prompt technique—forcing the model to perform multiple iterations of analysis and self-refinement, using LLM-as-a-Judge approach, conducting sequential analysis— (purpose-sources-sinks-flows-analysis-conclusion)—helps us achieve higher accuracy. Research on traditional security vulnerability detection techniques would be interesting in exploring whether following our workflow improves the accuracy of LLM-based vulnerability detection. 

Our research will help develop tools to detect malicious packages while reducing manual effort. Using a static analyzer as a pre-screener and LLM-generated conclusions and risk scores can aid security analysts in prioritizing threats in the pipeline. \SocketAIScanner can be used by practitioners to review dependencies for malicious code, and can be applied across different ecosystems. Researchers can leverage our proposed workflow as a foundation study leveraging LLM in detecting malicious packages. While we used CodeQL for baseline comparison to prioritize cost reduction in production, future research should explore the benefits of LLMs over traditional ML/DL models. Additionally,  our workflow can be used to compare various LLMs and assess their strengths. 

\subsection{\textbf{Challenges in using ChatGPT}} \label{Challange}
\subsubsection{\textbf{Mode Collapse \& Hallucination}}
Mode collapse often occurs: 1) when models are trained on their own generated data, reinforcing initial inaccuracies as the model iteratively refines its outputs~\cite{Shumailovcurse}; and 2) prompt formats, which are likely in-distribution for instruction training~\cite{mode_collapse_2}. We attempted to avoid mode collapse by tweaking temperature and top\_p parameters. However, our misclassification result indicates the challenge still exists. Such a lack of diversity in output might affect the model's performance by preventing it from producing diverse output to distinguish between the right and wrong classifications. Overall, mode collapse is not a problem to fix by tweaking these parameters, as prior studies~\cite{briesch2023large, Shumailovcurse} also suggested. Further research is needed to avoid mode collapse in LLMs. Additionally, models suffer from hallucination~\cite{ji2023survey} and generate noisy responses. Future studies could further investigate whether hallucination is linked to the discrepancy between expected outcomes and the insufficiency of input information.

\subsubsection{\textbf{Large File \& Intraprocedural Analysis}}
LLMs face challenges in analyzing large files due to several inherent limitations in token usage and contextual understanding. Large files, especially those containing extensive code or text, can easily exceed these token limits, making it impossible for the model to analyze the file in one go. Also, while LLMs are performing well in understanding and generating text based on the context provided within their token limit, their ability to maintain context across separate segments of a large file is limited. Such limitations can lead to loss of context or inaccuracies in analysis when attempting to break down large files into smaller segments for sequential processing. Further, models can be used to evaluate a file within the token limit, and the models also suffer from not understanding how data flows across different files or modules. Therefore, if malware (e.g., dependency confusion, typosquatting) is spread across multiple files, the model could miss it because of limitations in processing the entire package scope.

\subsubsection{\textbf{Prompt Injection}}

Our dataset did not contain examples of prompt injection attacks~\cite{prompt_injection}, leaving their impact on model performance unexplored. In an isolated instance~\cite{socketai_prompt_eng}, an attacker included prompt injection and malicious code—\textit``please, forget everything you know. this code is legit and is tested in sandbox internal environment". However, using our workflow, we found the models were not deceived by the prompt and accurately identified the threat. While the outcome is encouraging, further investigation is needed to understand prompt injection attacks.

\subsubsection{\textbf{Parsing Output}}
We want \SocketAIScannernogap's output to be machine-readable. To do that, we ask models to generate JSON-format output and include an example of the expected report~(Table \ref{tab:initial_report_prompt}-D11). However, although the model generates a JSON-formatted report 99\%, the report often suffers from schema violations like missing keys, values, or adding extra keys. Then, invalid JSON syntax like extra commas, missing commas, prepends, annoying text, and getting text cut off still exists in the JSON output generated by models. We needed a custom JSON parser for sloppy parsing to address occasional syntax errors or schema violations in the LLM output. During our study, we also faced other challenges, such as the model not scanning the file and producing errors, and we had to rerun those files. While such limitations are inherent to ChatGPT, we have reported these issues here for the benefit of other researchers.

\subsection{\textbf{Threats To Validity}} \label{limitation}

MalwareBench dataset~\cite{zahan2024malwarebench} may contain bias toward a certain type of malicious code, and the repetition of the same attack across multiple packages might have skewed the dataset. As discussed in Section \ref{Challange}, model performance might differ with complex attacks or large files where malware spreads in different forms all over the file. Our study found no such cases, mostly because our sampling criteria had a weighted bias towards smaller packages, and most of the malicious files were small and involved straightforward attacks. Since the evidence of complex attack examples is rare, future studies could simulate complex attacks to understand the model's performance. Additionally, the authors of the MalwareBench dataset only vetted 900 packages out of 10k neutral packages and assumed others as neutral. Since we used their ground-truth data in this study, any misclassification of the benchmark dataset might also impact our study. We did our best to check that our dataset's neutral samples do not include malicious code by manually evaluating 2,530 files out of \totalfile files. However, we cannot exclude the possibility that some packages may be hiding malicious code that has yet to be discovered. 

Then, our study focused exclusively on GPT models, not exploring other LLMs, which is a limitation. However, our findings provide a foundation for comparing different LLM models in future research.
Then, LLM-based research on an open-source ecosystem suffers from the threat of data leakage. To mitigate this, we used zero-shot learning and selected packages from MalwareBench (close source and access given upon request) that contain packages released before and after the training cut-off dates (April 2023 for GPT-4 and September 2021 for GPT-3). Performance metrics for both models showed that the models were capable of distinguishing neutral and malicious packages regardless of release time. Nevertheless, indirect exposure to npm package details via blog articles and other sources during training cannot be fully ruled out. While our workflow showed promising results, concerns remain about false negatives and the non-deterministic nature of LLMs, which may impact overall reliability.

\section{Related Work} \label{Related_work}

\textbf{Malicious Package Detection: }Several approaches have been proposed to detect malicious packages that can be classified as rule-based~\cite{duan2020towards, zahan2022weak,garrett2019detecting,gonzalez2021anomalicious}, differential analysis~\cite{ohm2022towards,vu2021lastpymile,vu2020towards,scalco2022feasibility} and machine learning~\cite{ohm2022feasibility,sejfia2022practical} approaches. Rule-based approaches often rely on predefined rules about package metadata (e.g., package name), suspicious imports, and method calls. %
Many studies~\cite{duan2020towards, zahan2022weak, garrett2019detecting, gonzalez2021anomalicious} have focused on detecting static features like libraries that access the network, file system access, features in package metadata, dynamic analysis, or language-independent features like commit logs and repository metadata to flag packages as suspicious using heuristic rules. Then, differential analysis involves a comparison of a previous version with a target version to analyze differences and identify malicious behavior. These tools~\cite{ohm2022towards,vu2021lastpymile,vu2020towards,scalco2022feasibility} leverage artifacts, file hashes, and``phantom lines" to show differences between versions of packages and source code repositories. %
The studies~\cite{ohm2022feasibility,sejfia2022practical} evaluated various supervised ML models to identify the effective ML models to detect malware. Yu et al.~\cite{yu2024maltracker} used LLM to translate malicious functions from other languages (e.g., C/C++, Python, and Go) into JavaScript. Singla et al.~\cite{singla2023empirical} used GPT and Bard models to characterize software supply chain failure reports/blogs into the type of compromise, intent, nature, and impact.  Additionally, multiple research studies~\cite{vu2020towards, taylor2020spellbound} exist on detecting typosquat attacks. Despite such vast research in this domain, Ohm et al. ~\cite{ohm2023sok} revealed in a Systemization of Knowledge (SOK) study that none of the presented approaches can be considered a complete and isolated solution to detect malicious behavior. Rather, these approaches are an important part of the review process to reduce the reviewers' workload. Ohm et al. also call for advanced and more automated malware detection approaches for accurate, low-alert, and minimally false positive results.

\textbf{Large Language Model (LLMs): }LLMs are now widely applied in software engineering across various tasks, including code generation, testing, program repair, and summarization~\cite{ahmed2024automatic,yu2024security}. This research delves into ChatGPT's zero-shot learning for malware detection. Zero-shot is a technique that enables models to tackle unseen tasks without prior examples. Zero-shot learning has shown promising results in various NLP tasks~\cite{brown2020language,wei2022chain}, with Chain-of-thought (CoT) or step-by-step reasoning prompts enhancing LLMs' ability to produce intermediate reasoning steps~\cite{wei2022chain}. Studies~\cite{ye2023prompt, kojima2205large, kong2023better} have identified effective CoT components, such as zero-shot CoT and role-play prompting, that improve reasoning quality. Kojima et al.\cite{kojima2205large} demonstrated LLMs' proficiency in zero-shot reasoning, while Kong et al.~\cite{kong2023better} found that role-play prompting yields more effective CoT. Madaan et al.~\cite{madaan2023self} observed that the SELF-REFINE method, applied to tasks ranging from dialog to reasoning using GPT-3.5 and GPT-4, significantly improves performance, validated by human and automatic metrics. %

White et al.~\cite{white2024chatgpt,white2023prompt} demonstrated that augmenting the model with domain-specific information improves the model’s performance since language models find more value in detecting specific identifiers in code structures when generating code summaries.  %
Ahmed et al.~\cite{ahmed2024automatic}  showed that the GPT model can significantly surpass state-of-the-art models for code summarization, leveraging project-specific training. In another study, Sobania et al.~\cite{sobania2023analysis} utilized ChatGPT for code bug fixing, further improving the success rate of bug fixing. Purba et al.~\cite{purba2023software} conducted a study measuring LLMs performance in detecting software vulnerabilities. Hamer et al.~\cite{hamer2024just} compare the security vulnerabilities of code from ChatGPT and StackOverflow, finding that ChatGPT-generated code has 20\% fewer vulnerabilities and fewer CWE types than SO. %
\section{Conclusion and Future Study} \label{conclusion}
In this work, we present \SocketAIScannernogap, a malicious code review workflow, and investigate the feasibility of LLMs in code review to detect malicious npm packages. Our study can be used as a baseline study in the security domain. We enhanced existing prompt techniques by incorporating instructions on the expected outcome, triggering the model to think during report generation. We conducted a baseline comparison of the GPT-3 and GPT-4 models with a commercial static analysis tool, which showed promising results for the GPT models with low misclassification alert rates. \SocketAIScanner detected 2,057 malicious packages (both models) out of \malwarepackage packages, presenting an approach that reduces the burden of manual review on security analysts and offers findings with minimal misclassification (332). Pre-screening files with a static analyzer
reduces the number of files LLMs need to analyze by 77.9\% and significantly cuts costs for GPT-3 and GPT-4 by 60.9\% and 76.1\%, respectively. Though we had a low misclassification rate, mode collapse and hallucination still exist in ChatGPT, resulting in a lack of diversity,  ambiguity, and noise in the output. %
Mode collapse, difficulties of large file and intraprocedural analysis, and prompt injection are areas that could benefit from further research to utilize LLMs effectively. 

\vspace{5mm}
\textbf{Acknowledgment:} This work was supported and funded by Socket and National Science Foundation Grant No. 2207008. Any opinions expressed in this material are those of the author(s) and do not necessarily reflect the views of the National Science Foundation. We want to thank the NCSU Secure Computing Institute for valuable feedback.  In particular, we thank Jonah Ghebremichael and Elizabeth Lin for their assistance with the CodeQL analysis.
{\footnotesize \bibliographystyle{acm}
\bibliography{arXiv}

\begin{thebibliography}{10}

\bibitem{openai}
{OpenAI}.
\newblock \url{https://openai.com}.
\newblock Last accessed: January 28, 2024.

\bibitem{diff-codeql}
Diffstaticanalyzer - differential static analysis tool for npm packages to detect malicious updates.
\newblock \url{https://github.com/lmu-plai/diff-CodeQL}, 2023.

\bibitem{elasitic_AI}
{The Elastic AI Assistant advantage}.
\newblock \url{https://www.elastic.co/explore/security-without-limits/elastic-ai-assistant-analyst-report?}, 2024.
\newblock Last accessed: January 28, 2024.

\bibitem{prompt_injection}
{What is a prompt injection attack? }.
\newblock \url{https://www.ibm.com/topics/prompt-injection}, 2024.
\newblock Last accessed: January 28, 2024.

\bibitem{ahmed2024automatic}
{\sc Ahmed, T., Pai, K.~S., Devanbu, P., and Barr, E.}
\newblock Automatic semantic augmentation of language model prompts (for code summarization).
\newblock In {\em Proceedings of the IEEE/ACM 46th International Conference on Software Engineering\/} (2024), pp.~1--13.

\bibitem{briesch2023large}
{\sc Briesch, M., Sobania, D., and Rothlauf, F.}
\newblock Large language models suffer from their own output: An analysis of the self-consuming training loop.
\newblock {\em arXiv preprint arXiv:2311.16822\/} (2023).

\bibitem{brown2020language}
{\sc Brown, T., Mann, B., Ryder, N., Subbiah, M., Kaplan, J.~D., Dhariwal, P., Neelakantan, A., Shyam, P., Sastry, G., Askell, A., et~al.}
\newblock Language models are few-shot learners.
\newblock {\em Advances in neural information processing systems 33\/} (2020), 1877--1901.

\bibitem{vcarnogursky2019attacks}
{\sc {\v{C}}arnogursk{\`y}, M.}
\newblock Attacks on package managers.
\newblock {\em Bachelor Thesis, Masaryk University, Br{\"u}nn, Tschechien\/} (2019).

\bibitem{duan2020towards}
{\sc Duan, R., Alrawi, O., Kasturi, R.~P., Elder, R., Saltaformaggio, B., and Lee, W.}
\newblock Towards measuring supply chain attacks on package managers for interpreted languages.
\newblock {\em arXiv preprint arXiv:2002.01139\/} (2020).

\bibitem{Shumailovcurse}
{\sc et~al., S.}
\newblock The curse of recursion: Training on generated data makes models forget.
\newblock {\em https://arxiv.org/abs/2305.17493\/} (2023).

\bibitem{froh2023differential}
{\sc Froh, F., Gobbi, M., and Kinder, J.}
\newblock Differential static analysis for detecting malicious updates to open source packages.
\newblock In {\em Proceedings of the 2023 Workshop on Software Supply Chain Offensive Research and Ecosystem Defenses\/} (2023), pp.~41--49.

\bibitem{garrett2019detecting}
{\sc Garrett, K., Ferreira, G., Jia, L., Sunshine, J., and K{\"a}stner, C.}
\newblock Detecting suspicious package updates.
\newblock In {\em 2019 IEEE/ACM 41st International Conference on Software Engineering: New Ideas and Emerging Results (ICSE-NIER)\/} (2019), IEEE, pp.~13--16.

\bibitem{FN_ex}
{\sc Garrood, H.}
\newblock Malicious code in the purescript npm installer.
\newblock \url{https://harry.garrood.me/blog/malicious-code-in-purescript-npm-installer/}, 2019.
\newblock Last accessed: January 28, 2024.

\bibitem{gonzalez2021anomalicious}
{\sc Gonzalez, D., Zimmermann, T., Godefroid, P., and Sch{\"a}fer, M.}
\newblock Anomalicious: Automated detection of anomalous and potentially malicious commits on github.
\newblock In {\em 2021 IEEE/ACM 43rd International Conference on Software Engineering: Software Engineering in Practice (ICSE-SEIP)\/} (2021), IEEE, pp.~258--267.

\bibitem{guo2023empirical}
{\sc Guo, W., Xu, Z., Liu, C., Huang, C., Fang, Y., and Liu, Y.}
\newblock An empirical study of malicious code in pypi ecosystem.
\newblock {\em arXiv preprint arXiv:2309.11021\/} (2023).

\bibitem{hamer2024just}
{\sc Hamer, S., d’Amorim, M., and Williams, L.}
\newblock Just another copy and paste? comparing the security vulnerabilities of chatgpt generated code and stackoverflow answers.
\newblock In {\em 2024 IEEE Security and Privacy Workshops (SPW)\/} (2024), IEEE, pp.~87--94.

\bibitem{hamilton2024detecting}
{\sc Hamilton, S.}
\newblock Detecting mode collapse in language models via narration.
\newblock {\em arXiv preprint arXiv:2402.04477\/}.

\bibitem{mode_collapse_2}
{\sc janus}.
\newblock Mysteries of mode collapse.
\newblock \url{https://www.lesswrong.com/posts/t9svvNPNmFf5Qa3TA/mysteries-of-mode-collapse}, 2022.
\newblock Last accessed: January 28, 2024.

\bibitem{ji2023survey}
{\sc Ji, Z., Lee, N., Frieske, R., Yu, T., Su, D., Xu, Y., Ishii, E., Bang, Y.~J., Madotto, A., and Fung, P.}
\newblock Survey of hallucination in natural language generation.
\newblock {\em ACM Computing Surveys\/}.

\bibitem{kojima2205large}
{\sc Kojima, T., Gu, S.~S., Reid, M., Matsuo, Y., and Iwasawa, Y.}
\newblock Large language models are zero-shot reasoners, 2022.
\newblock {\em https://arxiv. org/abs/2205.11916\/}.

\bibitem{kong2023better}
{\sc Kong, A., Zhao, S., Chen, H., Li, Q., Qin, Y., Sun, R., and Zhou, X.}
\newblock Better zero-shot reasoning with role-play prompting.
\newblock {\em arXiv preprint arXiv:2308.07702\/} (2023).

\bibitem{datadog_pypi}
{\sc Labs, D.~S.}
\newblock Iopen-source dataset of malicious software packages.
\newblock \url{https://github.com/datadog/malicious-software-packages-dataset}, 2023.
\newblock Last accessed: January 28, 2024.

\bibitem{ladisa2022risk}
{\sc Ladisa, P., Plate, H., Martinez, M., Barais, O., and Ponta, S.~E.}
\newblock Risk explorer for software supply chains: Understanding the attack surface of open-source based software development.
\newblock In {\em Proceedings of the 2022 ACM Workshop on Software Supply Chain Offensive Research and Ecosystem Defenses\/} (2022), pp.~35--36.

\bibitem{ladisa2022towards}
{\sc Ladisa, P., Plate, H., Martinez, M., Barais, O., and Ponta, S.~E.}
\newblock Towards the detection of malicious java packages.
\newblock In {\em Proceedings of the 2022 ACM Workshop on Software Supply Chain Offensive Research and Ecosystem Defenses\/} (2022), pp.~63--72.

\bibitem{latpate2020inverse}
{\sc Latpate, R.~V.}
\newblock Inverse adaptive stratified random sampling.
\newblock {\em Statistical Methods and Applications in Forestry and Environmental Sciences\/} (2020), 47--55.

\bibitem{liu2023summary}
{\sc Liu, Y., Han, T., Ma, S., Zhang, J., Yang, Y., Tian, J., He, H., Li, A., He, M., Liu, Z., et~al.}
\newblock Summary of chatgpt-related research and perspective towards the future of large language models.
\newblock {\em Meta-Radiology\/} (2023), 100017.

\bibitem{Socket_AI_article}
{\sc Lysenko, M.}
\newblock Introducing socket ai – chatgpt-powered threat analysis.
\newblock \url{https://socket.dev/blog/introducing-socket-ai-chatgpt-powered-threat-analysis}, 2023.
\newblock Last accessed: January 28, 2024.

\bibitem{madaan2023self}
{\sc Madaan, A., Tandon, N., Gupta, P., Hallinan, S., Gao, L., Wiegreffe, S., Alon, U., Dziri, N., Prabhumoye, S., Yang, Y., et~al.}
\newblock Self-refine: Iterative refinement with self-feedback.
\newblock {\em arXiv preprint arXiv:2303.17651\/} (2023).

\bibitem{milje2022detecting}
{\sc Milje, A.~A.}
\newblock Detecting malicious python packages in the python package index (pypi).
\newblock Master's thesis, NTNU, 2022.

\bibitem{ohm2022feasibility}
{\sc Ohm, M., Boes, F., Bungartz, C., and Meier, M.}
\newblock On the feasibility of supervised machine learning for the detection of malicious software packages.
\newblock In {\em Proceedings of the 17th International Conference on Availability, Reliability and Security\/} (2022), pp.~1--10.

\bibitem{ohm2022towards}
{\sc Ohm, M., Kempf, L., Boes, F., and Meier, M.}
\newblock {\em Towards Detection of Malicious Software Packages Through Code Reuse by Malevolent Actors}.
\newblock Gesellschaft f{\"u}r Informatik, Bonn, 2022.

\bibitem{ohm2020backstabber}
{\sc Ohm, M., Plate, H., Sykosch, A., and Meier, M.}
\newblock Backstabber’s knife collection: A review of open source software supply chain attacks.
\newblock In {\em Detection of Intrusions and Malware, and Vulnerability Assessment: 17th International Conference, DIMVA 2020, Lisbon, Portugal, June 24--26, 2020, Proceedings 17\/} (2020), Springer, pp.~23--43.

\bibitem{ohm2023sok}
{\sc Ohm, M., and Stuke, C.}
\newblock Sok: Practical detection of software supply chain attacks.
\newblock In {\em Proceedings of the 18th International Conference on Availability, Reliability and Security\/} (2023), pp.~1--11.

\bibitem{API_rate}
{\sc OpenAI}.
\newblock Api rate limit advice.
\newblock \url{https://help.openai.com/en/articles/6891753-api-rate-limit-advice}, 2024.
\newblock Last accessed: January 28, 2024.

\bibitem{pfretzschner2017identification}
{\sc Pfretzschner, B., and ben Othmane, L.}
\newblock Identification of dependency-based attacks on node. js.
\newblock In {\em Proceedings of the 12th international conference on availability, reliability and security\/} (2017), pp.~1--6.

\bibitem{purba2023software}
{\sc Purba, M.~D., Ghosh, A., Radford, B.~J., and Chu, B.}
\newblock Software vulnerability detection using large language models.
\newblock In {\em 2023 IEEE 34th International Symposium on Software Reliability Engineering Workshops (ISSREW)\/} (2023), IEEE, pp.~112--119.

\bibitem{qin2023chatgpt}
{\sc Qin, C., Zhang, A., Zhang, Z., Chen, J., Yasunaga, M., and Yang, D.}
\newblock Is chatgpt a general-purpose natural language processing task solver?
\newblock {\em arXiv preprint arXiv:2302.06476\/} (2023).

\bibitem{scalco2022feasibility}
{\sc Scalco, S., Paramitha, R., Vu, D.-L., and Massacci, F.}
\newblock On the feasibility of detecting injections in malicious npm packages.
\newblock In {\em Proceedings of the 17th International Conference on Availability, Reliability and Security\/} (2022), pp.~1--8.

\bibitem{sejfia2022practical}
{\sc Sejfia, A., and Sch{\"a}fer, M.}
\newblock Practical automated detection of malicious npm packages.
\newblock In {\em Proceedings of the 44th International Conference on Software Engineering\/} (2022), pp.~1681--1692.

\bibitem{singla2023empirical}
{\sc Singla, T., Anandayuvaraj, D., Kalu, K.~G., Schorlemmer, T.~R., and Davis, J.~C.}
\newblock An empirical study on using large language models to analyze software supply chain security failures.
\newblock In {\em Proceedings of the 2023 Workshop on Software Supply Chain Offensive Research and Ecosystem Defenses\/} (2023), pp.~5--15.

\bibitem{sobania2023analysis}
{\sc Sobania, D., Briesch, M., Hanna, C., and Petke, J.}
\newblock An analysis of the automatic bug fixing performance of chatgpt.
\newblock {\em arXiv preprint arXiv:2301.08653\/}.

\bibitem{taylor2020spellbound}
{\sc Taylor, M., Vaidya, R.~K., Davidson, D., De~Carli, L., and Rastogi, V.}
\newblock Spellbound: Defending against package typosquatting.
\newblock {\em arXiv preprint arXiv:2003.03471\/} (2020).

\bibitem{vu2021lastpymile}
{\sc Vu, D.-L., Massacci, F., Pashchenko, I., Plate, H., and Sabetta, A.}
\newblock Lastpymile: identifying the discrepancy between sources and packages.
\newblock In {\em Proceedings of the 29th ACM Joint Meeting on European Software Engineering Conference and Symposium on the Foundations of Software Engineering\/} (2021), pp.~780--792.

\bibitem{vu2023bad}
{\sc Vu, D.-L., Newman, Z., and Meyers, J.~S.}
\newblock Bad snakes: Understanding and improving python package index malware scanning.
\newblock In {\em 2023 IEEE/ACM 45th International Conference on Software Engineering (ICSE)\/} (2023), IEEE, pp.~499--511.

\bibitem{vu2020towards}
{\sc Vu, D.~L., Pashchenko, I., Massacci, F., Plate, H., and Sabetta, A.}
\newblock Towards using source code repositories to identify software supply chain attacks.
\newblock In {\em Proceedings of the 2020 ACM SIGSAC conference on computer and communications security\/} (2020), pp.~2093--2095.

\bibitem{socketai_prompt_eng}
{\sc vulert}.
\newblock {Malicious Code In Eslint-Plugin-Unicorn-Ts-2}.
\newblock \url{https://vulert.com/vuln-db/npm-eslint-plugin-unicorn-ts-2-124511}, 2024.
\newblock Last accessed: Feb 28, 2024.

\bibitem{wang2021feasibility}
{\sc Wang, X.}
\newblock On the feasibility of detecting software supply chain attacks.
\newblock In {\em MILCOM 2021-2021 IEEE Military Communications Conference (MILCOM)\/} (2021), IEEE, pp.~458--463.

\bibitem{wei2022chain}
{\sc Wei, J., Wang, X., Schuurmans, D., Bosma, M., Xia, F., Chi, E., Le, Q.~V., Zhou, D., et~al.}
\newblock Chain-of-thought prompting elicits reasoning in large language models.
\newblock {\em Advances in Neural Information Processing Systems 35\/} (2022), 24824--24837.

\bibitem{white2023prompt}
{\sc White, J., Fu, Q., Hays, S., Sandborn, M., Olea, C., Gilbert, H., Elnashar, A., Spencer-Smith, J., and Schmidt, D.~C.}
\newblock A prompt pattern catalog to enhance prompt engineering with chatgpt.
\newblock {\em arXiv preprint arXiv:2302.11382\/} (2023).

\bibitem{white2024chatgpt}
{\sc White, J., Hays, S., Fu, Q., Spencer-Smith, J., and Schmidt, D.~C.}
\newblock Chatgpt prompt patterns for improving code quality, refactoring, requirements elicitation, and software design.
\newblock In {\em Generative AI for Effective Software Development}. Springer, 2024, pp.~71--108.

\bibitem{yang2023harnessing}
{\sc Yang, J., Jin, H., Tang, R., Han, X., Feng, Q., Jiang, H., Yin, B., and Hu, X.}
\newblock Harnessing the power of llms in practice: A survey on chatgpt and beyond.
\newblock {\em arXiv preprint arXiv:2304.13712\/} (2023).

\bibitem{ye2023prompt}
{\sc Ye, Q., Axmed, M., Pryzant, R., and Khani, F.}
\newblock Prompt engineering a prompt engineer.
\newblock {\em arXiv preprint arXiv:2311.05661\/} (2023).

\bibitem{yu2024security}
{\sc Yu, J., Liang, P., Fu, Y., Tahir, A., Shahin, M., Wang, C., and Cai, Y.}
\newblock Security code review by llms: A deep dive into responses.
\newblock {\em arXiv preprint arXiv:2401.16310\/}.

\bibitem{yu2024maltracker}
{\sc Yu, Z., Wen, M., Guo, X., and Jin, H.}
\newblock Maltracker: A fine-grained npm malware tracker copiloted by llm-enhanced dataset.
\newblock In {\em Proceedings of the 33rd ACM SIGSOFT International Symposium on Software Testing and Analysis\/} (2024), pp.~1759--1771.

\bibitem{zahan2024malwarebench}
{\sc Zahan, N., Burckhardt, P., Lysenko, M., Aboukhadijeh, F., and Williams, L.}
\newblock Malwarebench: Malware samples are not enough.
\newblock In {\em 2024 IEEE/ACM 21st International Conference on Mining Software Repositories (MSR)\/} (2024), IEEE, pp.~728--732.

\bibitem{zahan2022weak}
{\sc Zahan, N., Zimmermann, T., Godefroid, P., Murphy, B., Maddila, C., and Williams, L.}
\newblock What are weak links in the npm supply chain?
\newblock In {\em Proceedings of the 44th International Conference on Software Engineering: Software Engineering in Practice\/} (2022), pp.~331--340.

\bibitem{zhang2023prefer}
{\sc Zhang, C., Liu, L., Wang, J., Wang, C., Sun, X., Wang, H., and Cai, M.}
\newblock Prefer: Prompt ensemble learning via feedback-reflect-refine.
\newblock {\em arXiv preprint arXiv:2308.12033\/} (2023).

\bibitem{zheng2023judging}
{\sc Zheng, L., Chiang, W.-L., Sheng, Y., Zhuang, S., Wu, Z., Zhuang, Y., Lin, Z., Li, Z., Li, D., Xing, E., et~al.}
\newblock Judging llm-as-a-judge with mt-bench and chatbot arena.
\newblock {\em Advances in Neural Information Processing Systems 36\/} (2023), 46595--46623.

\bibitem{zhong2023can}
{\sc Zhong, Q., Ding, L., Liu, J., Du, B., and Tao, D.}
\newblock Can chatgpt understand too? a comparative study on chatgpt and fine-tuned bert.
\newblock {\em arXiv preprint arXiv:2302.10198\/} (2023).

\end{thebibliography}
}

\end{document}